\documentclass[twocolumn, pra, aps, amsfonts, amsmath, amssymb]{revtex4-1} 

\usepackage{graphicx,bbm,enumerate,float}
\usepackage[colorlinks=true,bookmarks=false,citecolor=blue,urlcolor=blue]{hyperref}

\graphicspath{ {./images/} }
\usepackage[utf8]{inputenc}
\usepackage[export]{adjustbox}
\usepackage{physics}
\usepackage{soul}
\usepackage{comment}
\usepackage{svg}
\usepackage{subfigure}
\usepackage{cleveref}
\crefname{equation}{Eq}{Eqs} 
\crefrangelabelformat{equation}{(#3#1#4--#5#2#6)}

\everymath{\displaystyle}

\newcommand{\orcid}[1]{\href{https://orcid.org/#1}{\includegraphics[width=10pt]{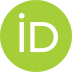}}}

\newcommand{\beq}{\begin{eqnarray}}
\newcommand{\eeq}{\end{eqnarray}}
\begin{document}

\title{Entanglement dynamics via Geometric phases in Trapped-ions}

\author{Dharmaraj Ramachandran\orcid{0000-0002-3068-1586}}
\email{p20200040@goa.bits-pilani.ac.in}
\author{Ganesh Hanchanahal\orcid{0009-0004-6531-8009}}
\email{f20200571@goa.bits-pilani.ac.in}
\author{Radhika Vathsan\orcid{0000-0001-5892-9275}}
\email{radhika@goa.bits-pilani.ac.in}

\affiliation{Physics Department,\\ BITS-Pilani, K. K. Birla Goa Campus, Goa 403726, INDIA}

\date{\today}
\begin{abstract}
Trapped-ion systems are a leading platform for quantum computing. The M{\o}lmer-S{\o}rensen (MS) gate is a widely used method for implementing controlled interactions in multipartite systems. However, due to unavoidable interactions with the environment, quantum states undergo non-unitary evolution, leading to significant deviations from ideal dynamics.

Common techniques such as Quantum Process Tomography (QPT) and Bell State Tomography (BST) are typically employed to evaluate MS gate performance and to characterize noise in the system. In this work, we propose leveraging the geometric phase(GP) as a tool for performance assessment and noise identification in the MS gate. Our findings indicate that the GP is particularly sensitive to environmental noise occurring around twice the clock pulse time. Given that GP measurements do not require full-state tomography, this approach offers a practical and experimentally feasible method to detect entanglement and classify the nature of noise affecting the system.

\end{abstract}
\maketitle

\section{Introduction}
A bipartite quantum system in the state \( \rho_{AB} \) in a composite Hilbert space \( H_A \otimes H_B \) is said to be entangled if and only if it cannot be expressed as a convex combination of product states: 
\begin{equation}
    \rho_{AB} \neq \sum_i p_i \rho_A^{(i)} \otimes \rho_B^{(i)},
\end{equation}
where \( p_i \) are probabilities, and \( \rho_A^{(i)} \) and \( \rho_B^{(i)} \) are density operators of subsystems \( A \) and \( B \), respectively.

Entanglement serves as a fundamental resource in numerous quantum information processing tasks, such as quantum teleportation and superdense coding~\cite{ent_horo_2009}. Beyond its wide-ranging applications in quantum computing and communication, entanglement represents a fundamental feature of quantum mechanics. Consequently, the ability to generate and preserve entanglement in quantum systems is essential to leverage its advantages effectively. 

Ions trapped inside an RF voltage trap---commonly known as a Paul trap---have been extensively used as a platform for generating entangled quantum states. Trapped ion systems are promising candidates as qubits for a universal quantum computer, particularly due to their long coherence times. Two well-known gates used for entangling trapped ions are the Cirac-Zoller (CZ) gate~\cite{ExptG40} and the M{\o}lmer-S{\o}rensen (MS) gate~\cite{ExptG38, ExptG39}.  
While the CZ gate can create entangled ions, it requires ground-state cooling of the ions, which is experimentally challenging. This, in turn, has made the MS gate a preferred scheme for generating entangled states in recent times. 

As with any quantum system, trapped ions are susceptible to environmental noise. Such noise interacts with the entangling gate operation, leading to non-trivial trajectories of the quantum state in parameter space. Identifying the nature of these noises and understanding their corresponding trajectories is crucial for effective noise suppression and mitigation.

The standard method to analyze the performance of the MS gate and calculate associated noise is to create a Bell state and perform partial state tomography~\cite{Bst_expt1,Bst_expt2, Bst_expt3, thesis}. This method is also known as Bell State Tomography (BST) where the fidelity between the quantum state and a target bell state is estimated.

A more detailed approach is Quantum Process Tomography (QPT), through which a process matrix can be evaluated. This matrix fully captures unitary dynamics, noise and systematic errors for a quantum process, subject to assumptions regarding  the  memory of the environment. H.~N.~Tinkey \textit{et al.} have performed a detailed study of QPT for the MS gate~\cite{qpt_ms}.

While BST and QPT are comprehensive techniques to analyze gate performance, they require multiple measurements of the quantum state. This requirement appears to be even more challenging as the number of measurements grows exponentially with system size. 

In this article, we propose utilizing the Geometric Phase (GP) as an effective tool to analyze the performance of the MS gate. We refer to this method as Geometric Phase-based Performance Analysis (GPA). 

The dynamical part of GP is a global phase acquired by a quantum state during non-parallel transport in parameter space. As a single parameter, GP acts as an effective memory of the quantum state, encoding information about its evolution.

 GPA is not an entirely new concept and several authors have explored similar ideas in different contexts. Some studies have proposed using the GP as an indicator of qubit-environment coupling in single-qubit systems and have suggested parameter values for which noise-induced corrections to the GP are more pronounced~\cite{gpa1, Product6, product13, srikanth_qubit}. In the case of non-Markovian interactions, X.~X.~Yi \textit{et al.} have derived analytical expressions for the GP when a single qubit is subjected to a non-Markovian dephasing environment~\cite{Product4}, while J.~J.~Chen \textit{et al.} have investigated conditions under which the non-Markovianity of system-environment coupling becomes more pronounced in single-qubit systems~\cite{product11}. 

For entangled systems, a few studies have examined the GP acquired due to the unitary dynamics of initially entangled states~\cite{entangled6, tavis_cumming, exptG3, exptG2, exptG1}, demonstrating that the presence or absence of entanglement can often be inferred from the behavior of the GP in those systems. In the case of trapped-ion systems, L.~U.~Hong-Xia has studied the GP arising in unitary dynamics, highlighting the effects of the initial state on the acquired GP~\cite{ion_gp_2}, while K.~M{\"u}ller \textit{et al.} have investigated GP in non-unitary dynamics, proposing strategies to mitigate environment-induced modifications to the GP~\cite{ion_gp1}. However, to the best of our knowledge, no study has systematically explored the behavior of the GP during the implementation of the MS gate or its potential application in gate performance analysis. 

To implement GPA, one must measure the GP in an experimental setup. In trapped-ion systems, GP is typically measured using setups such as Ramsey interferometers. L.~U.~Hong-Xia \textit{et al.} have proposed a scheme for experimentally measuring GP in such systems~\cite{ion_gp_2}. Furthermore, experimentalists have successfully observed the GP in superconducting qubits and investigated its sensitivity to noise~\cite{gp_qubit_expt1, gp_qubit_expt2, gp_qubit_expt3}.

While GPA may not be as comprehensive as QPT or BST, it offers a feasible test to assess the strength and nature of specific noise sources and behaviors, particularly considering the fact that the quantity being calculated is always from a single parameter and does not scale with system size. 

The article is organized as follows. First, we introduce the working of the MS gate under both weak-field and strong-field conditions and discuss the kinematic approach to calculating the GP before presenting our findings. We then examine the behavior of GP under these conditions and demonstrate its utility in verifying the successful implementation of the MS gate in the weak-field regime. Subsequently, we discuss typical noise sources in such systems, highlighting how GP exhibits heightened sensitivity to external noise at a characteristic time interval of approximately twice the gate pulse time ($2T$). Finally, we explore the types of noise that induce non-trivial GP in subsystems before concluding the article.

\section{The MS Gate}

The MS gate was introduced by A.~S\o{}rensen and K.~M{\o}lmer~\cite{ExptG39,ExptG38}. While  we summarize its key results here,  we do recommend the following resources~\cite{thesis,duke_lectnotes} for a more detailed explanation.

Consider an ionic species with selected qubit levels denoted by \(\ket{0}\) and \(\ket{1}\), confined in a Paul trap. The transverse vibrational modes of the ions in the  trap are denoted by mode numbers \(n\). Gates are accomplished by coupling the ionic levels to the vibrational modes through optical pulses. The coupling between the incident light pulse and motional modes is quantified using the well known Lamb-Dicke parameter $\eta$. 
 For an ion in $n^{\text{th}}$ vibrational mode of frequency \(\nu\), the $(n+1)^{\text{th}}$ and the  $(n-1)^{\text{th}}$ modes are called the  blue sideband (BSB) and the red sideband (RSB) respectively.
 \begin{figure}[b]
    \centering
    \includegraphics[width=1\linewidth]{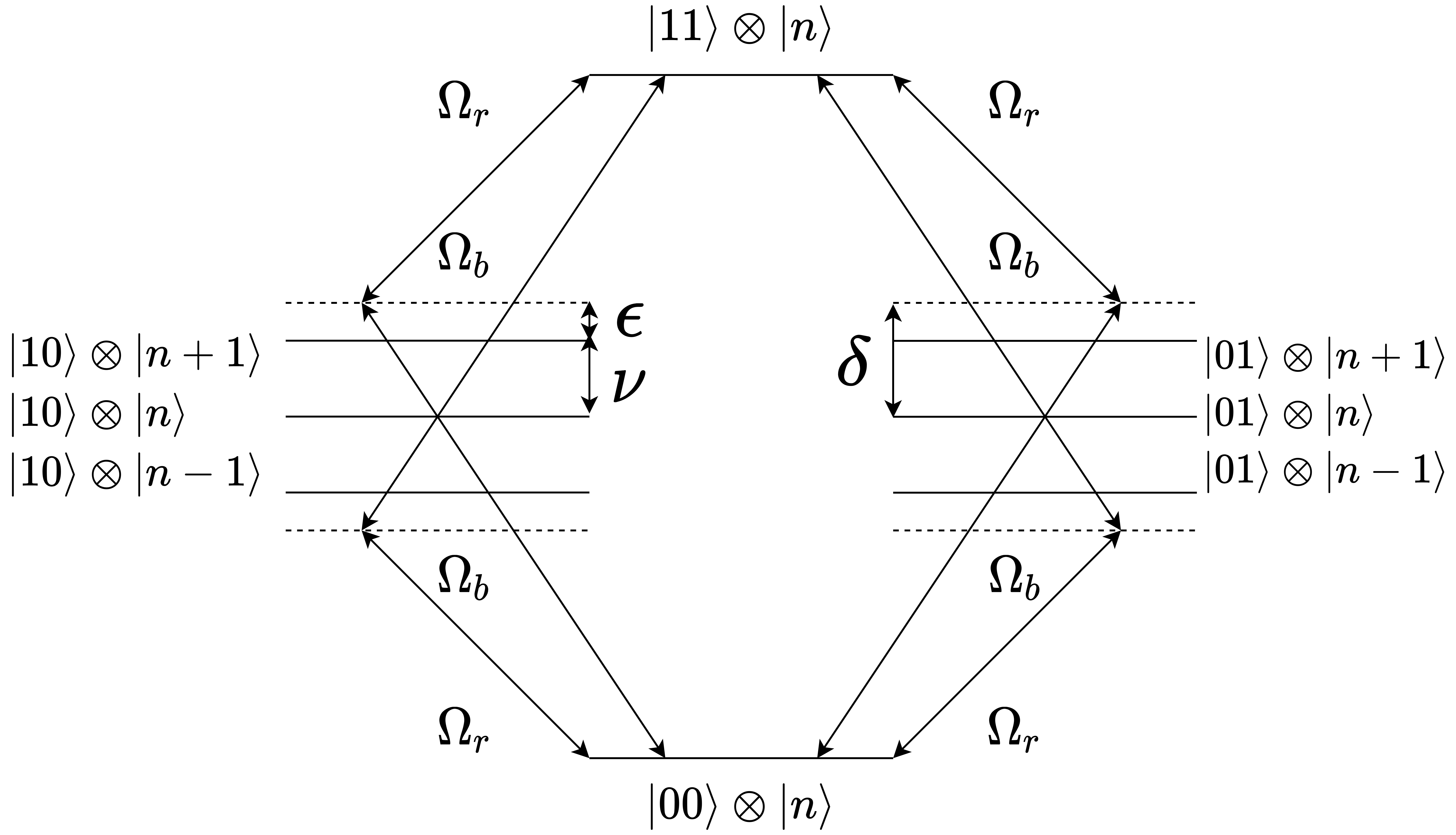}
    \caption{Energy band diagram for a two-qubit system subject to MS-interaction.}
    \label{Enlvl}
\end{figure}
When bichromatic light detuned by \(\delta = \nu + \epsilon\) to the blue and red sidebands is shone on each of two  ions, it drives  them simultaneously to RSB and BSB with phases $\phi_r$ and $\phi_b$ respectively. Defining the  spin phase as $\phi_s = \frac{1}{2}(\phi_b + \phi_r)$ and the  motional phase as   $\phi_m = \frac{1}{2}(\phi_b - \phi_r) $, the interaction Hamiltonian for each  ion is given by
\begin{equation}
      H_I(t)=\frac{\eta \hbar \Omega}{2} \sigma_\phi\left(a \mathrm{e}^{\mathrm{i} \epsilon t} \mathrm{e}^{\mathrm{i} \phi_m}+a^{\dagger} \mathrm{e}^{-\mathrm{i} \epsilon t} \mathrm{e}^{-\mathrm{i} \phi_m}\right)
\end{equation}
where $\sigma_{\phi} = \sigma_-e^{i\phi_s} - \sigma_+e^{-i\phi_s} $ and $\epsilon = \delta -\nu$. $ \Omega$ is the Rabi frequency. $ \sigma_{\pm} $ represent internal qubit raising and lowering operator and $a$ represents annihilation operator for the vibrational modes. It is to be noted that under Lamb Dicke limit $\Omega_b \approx \Omega_r \approx \Omega$, where $\Omega_{r/b}$ represent the Rabi frequency of RSB and BSB respectively. The two-ion  Hamiltonian is obtained by replacing $\sigma_\phi$ with the following spin operator:
\begin{equation}
    S= S_{-}e^{\mathrm{i}\phi_s} - S_{+}e^{-\mathrm{i}\phi_s},
\end{equation}
where $S_{+}$ and $S_{-}$ are defined as
\begin{align}
         S_{\pm}&=\sigma_{\pm}^{(1)} \otimes I^{(2)}+\hat{I}^{(1)} \otimes \sigma_{\pm}^{(2)}. 
\end{align}
The corresponding  unitary evolution operator for the two ions  is:
\begin{align}
    U_{\text{MS}}(t)  =& \mathrm{e}^{[S\left(\alpha(t) a+\alpha^*(t) a^{\dagger}\right) + \mathrm{i} S^2\beta(t)]} \nonumber \\
 =& D\big(\alpha(t) S\big) \mathrm{e}^{\mathrm{i}S^2\beta(t)}
   \label{Eq:unitary}
\end{align}
where the variables in the expression are defined as 
\begin{align}
    \alpha(t) &=\frac{\eta \Omega}{\epsilon} \mathrm{e}^{\mathrm{i} \epsilon t / 2} \sin 
                          (\frac{\epsilon t}{2}) \mathrm{e}^{\mathrm{i} \phi_m}, \\
     \beta(t) &= \left(\frac{\eta \Omega}{2 \epsilon}\right)^2(\epsilon t-\sin \epsilon t). 
\end{align}
The energy levels involved in  the MS interaction are shown in Fig:\ref{Enlvl}.

\noindent\textbf{Generation of  Entangled states}\\
The MS gate is typically operated under two regimes in order to generate maximally entangled states, namely the Weak Field (WF) regime and the Strong Field (SF) regime.

\begin{figure}[h!]
    \centering
    \includegraphics[width=1\linewidth]{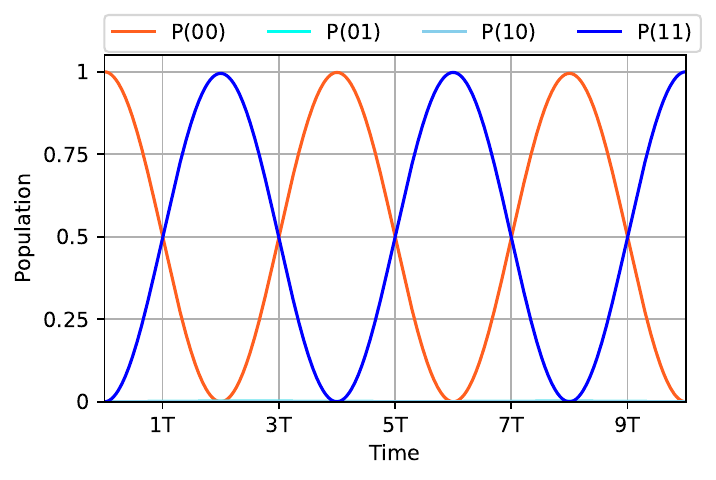}
    \caption{Population density of energy levels during MS interaction under WF condition:  $\eta = 0.1$, $\Omega = 0.1\nu$,  $\delta = 0.9 \nu$, as given in \cite{ExptG38}.}
    \label{fig:wf}
\end{figure}

\noindent \textbf{WF condition}\\
WF approximation implies that the Lamb-Dicke parameter($\eta$) is very low such that single photon excitations are completely suppressed permitting only two-photon processes satisfying the condition

\begin{equation}
    \eta\Omega \ll  \delta - \nu
\end{equation}
In this limit $\alpha(t)$ and $\beta(t)$ simplifies to
\begin{align}
    \alpha(t) &\approx 0  \quad \text{and}  \quad  
    \beta(t) \approx  \frac{\tilde{\Omega}t}{4}
    \label{eq:alp_bet_simpl}
\end{align}
where the effective Rabi frequency $\tilde{\Omega} = \frac{(\eta\Omega)}{\epsilon}^{2}$. 
Substituting Eq:~\eqref{eq:alp_bet_simpl} in Eq:~\eqref{Eq:unitary}
\begin{equation}
    U_{MS}(t)  = \mathrm{e}^{\mathrm{i}S^2\beta(t)}
\end{equation}
The action of this unitary leads to the following dynamics,
\begin{align}
     U_{MS}(t)\ket{00} = \cos(\frac{\tilde{\Omega}t}{2})\ket{00} &- \mathrm{i}\mathrm{e}^{-2\mathrm{i}\phi_s}\sin(\frac{\tilde{\Omega}t}{2})\ket{11}. 
\end{align}
This unitary leads to a maximally entangled state at $t =$ integer multiples of $  T =  \pi/2\tilde{\Omega}$
The diagonal entries of the two-qubit density operator that represent the population densities of each energy level have been plotted in Fig:~\ref{fig:wf}.\\

\begin{figure}[h!]
    \centering
\includegraphics[width=1\linewidth]{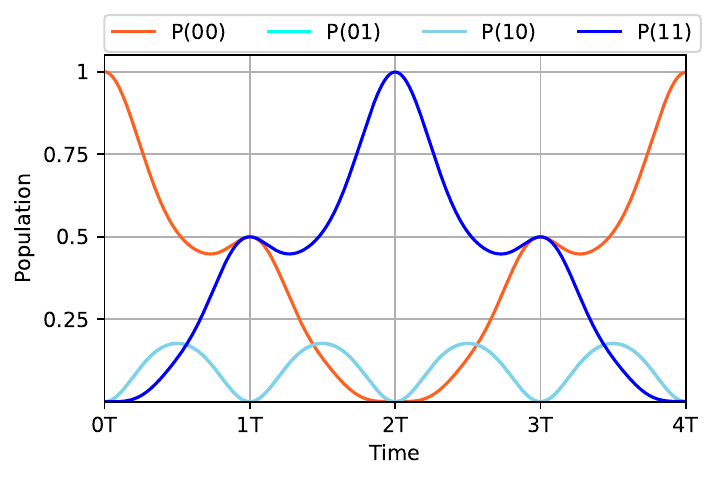}
    \caption{Population density of different energy levels when MS interaction is operated in SF regime: $\omega_m = 2\pi \times2.03 \text{MHz}$, $\delta = 2\pi \times 16.7\text{kHz}$, $\eta = 0.028$, $\Omega = 2\pi \times 270\text{kHz}$ which results in $T = 66\mu s$ as given in \cite{thesis}.}
    \label{fig:ff}
\end{figure}

\noindent \textbf{SF condition}\\
Under the SF condition, single-photon excitations are allowed to populate intermediate energy levels $\ket{10}$ and $\ket{01}$ while ensuring that this population becomes zero at gate pulse times ($T$) while fixing the detuning to a particular value as, 
\begin{align}
    \epsilon &= 2\Omega\eta \;\;\; \text{and} \nonumber \\
    T &= \frac{2\pi}{\epsilon} \label{eq:detun_condtn} .
\end{align}
By substituting Eq:~\eqref{eq:detun_condtn} in Eq:~\eqref{Eq:unitary} the unitary operator at gate pulse time ($T$) is written as 
\begin{equation}
    U(T)=\mathrm{e}^{-\mathrm{i} \frac{\pi}{8} S^2}
\end{equation}
For two ions in a separable state in the computational basis, the MS interaction forms a maximally entangled state at T as, 
\begin{align}
& U(t = T)|00\rangle=\frac{1}{\sqrt{2}}\left(|00\rangle+\mathrm{ie}^{2 \mathrm{i} \phi_s}|11\rangle\right) 
\end{align}
The population density of different energy levels is plotted in Fig: \ref{fig:ff}. 
While the intermediate energy levels $\ket{01}$ and $\ket{10}$ are populated they become zero at gate pulse times enabling maximal entanglement. 

\begin{figure}[h!]
    \centering
    \includegraphics[width=1\linewidth]{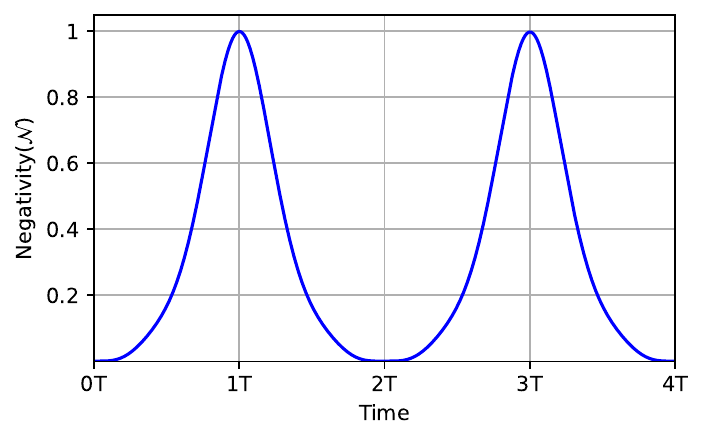}
    \caption{Oscillation of entanglement measured using negativity of quantum state initially in $\ket{00}$ state during MS interaction in SF regime }
    \label{fig:ff_ent}
\end{figure}

A well-known entanglement measure of a quantum state $\rho$ is negativity which is defined as \cite{negativity}, 
\begin{equation}
    \mathcal{N}(\rho) := \sum_{a_i<0} a_i 
\end{equation}
where $a_i$'s are the eigenvalues of the partial transposed density matrix. 
The oscillation of $\mathcal{N}$ during the MS interaction in the SF refime has been plotted in Fig: \ref{fig:ff_ent}. 

\section{GP: The kinematic approach}
When a quantum system undergoes an evolution, it might pick up a global phase based on the trajectory of the system in parameter space known as the GP. These GPs were first experimentally detected independently by Pancharatnam~\cite{ExptG20} and L. Higgins \textit{et al.} in 1950's~\cite{gp_first}. Later Berry\cite{ExptG21} provided a generalized formalism for GP arising in systems undergoing adiabatic, slow cyclic evolutions. A very important contribution for generalisng the formulation of GP was given by J. Samuel and R. Bhandari\cite{ExptG36}. 
In this work, we use the formulation given by N. Mukunda and R. Simon known as the kinematic approach for calculating the GP\cite{Mukunda1993}. It states that the GP picked up by a pure quantum state $\ket{\psi(t)}$ during the time interval $t \in [0, \tau]$ is given by 
\begin{align}
\phi_g(\tau) &= \phi_{\text{global}} - \phi_{\text{dynamical}} \nonumber \\
& = \arg(\bra{\psi(0)}\ket{\psi(\tau)}) - \text{Im} \int_0^{\tau} \bra{\psi(0)}\ket{\dot{\psi(t)}}dt
\end{align}
where $\dot{\psi(t)}$ represents $\frac{d }{dt}\psi(t)$. $\phi_{\text{global}}$ depends on only the initial and final states while $\phi_{\text{dynamical}}$ depends on the curvature of path in parameter space.   
Later this expression was generalised by D.M. Tong \textit{et al} \cite{ExptG26a} to mixed  states given by, 
\begin{equation}
\begin{aligned}
    \phi_g(\tau) = \text{Arg}\Bigg[ 
        \sum_k \sqrt{\epsilon_k(0)\epsilon_k(\tau)} 
        \langle\Psi_k(0)|\Psi_k(\tau)\rangle \\ 
        e^{-\int_0^\tau dt\langle\Psi_k(t)|\dot{\Psi}_k(t)\rangle} 
    \Bigg]
\end{aligned}
\end{equation}
where $|\Psi_k(t)\rangle$ are the instantaneous eigenstates of the density matrix and $\epsilon_k(t)$ are the instantaneous eigenvalues.
 When the system is initially a pure state given by $\rho(0) = \ket{\Psi}\bra{\Psi}$ the above expression reduces to~\cite{ExptG27}
\begin{align}
    \phi_g(\tau) = \text{Arg} \left\{\left\langle\Psi(0) \mid \Psi(\tau)\right\rangle\right\}\nonumber\\-\operatorname{Im} \int_0^\tau dt \left\langle\Psi\left(t\right) \mid \dot{\Psi}\left(t\right)\right\rangle
    \label{gp_expr}
\end{align}

Since MS gate is generally applied after a quantum system is prepared in an initial pure state, in all the simulations in this letter we have used Eq. \eqref{gp_expr} to calculate the GP considering the initial state to be $\ket{00}$. 

\section{GP under Unitary dynamics}
\noindent \textbf{WF regime}\\
As mentioned in Eq:~\eqref{wf_psi}, the quantum state of qubits in the WF regime of MS gate for an initial state $\ket{00}$ is given by,
\begin{equation}
    \ket{\psi(t)} = \cos(\frac{\tilde{\Omega} t}{2})\ket{00} + \sin(\frac{\tilde{\Omega} t}{2})\ket{11}.
    \label{wf_psi}
\end{equation}  

By substituting Eq:~\eqref{wf_psi} into Eq:~\eqref{gp_expr}, we see that the GP for an ideal WF condition is:  
\begin{equation}
    \phi_{wf}(t) = 0 \quad \forall t. 
\end{equation}  
This result implies that the state vector under the M{\o}lmer-S{\o}rensen (MS) interaction in the ideal WF regime undergoes parallel transport in parameter space.  

\begin{figure}[t]
    \centering
    \includegraphics[width=1\linewidth]{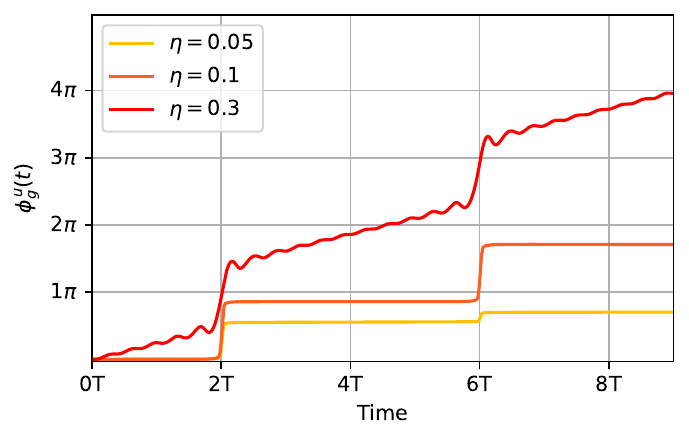}
    \caption{GP acquired by the quantum state undergoing MS-interaction under WF condition for different values of $\eta$. }
    \label{fig:wf_inaccuracy}
\end{figure}

However, as the Lamb-Dicke parameter increases, a non-trivial GP emerges in the WF regime, as depicted in Fig:~\ref{fig:wf_inaccuracy}.
As the value of $\eta$ increases, the dynamical phase acquired during the evolution increases significantly. 
This observation suggests that the GP acquired by the quantum state in the WF condition can serve as a diagnostic tool to evaluate the extent to which the WF condition has been achieved.  
Even though the MS gate under WF interaction is insensitive to changes in motional degrees of freedom it has a major drawback of high gate pulse time. A typical gate pulse time in the WF regime implemented on $\text{Ca}^+$ ions is roughly 5-10 times higher than the SF regime. Due to this limitation, MS gates are typically operated in SF regimes.

\begin{figure}[!h]
    \centering
    \includegraphics[width=1\linewidth]{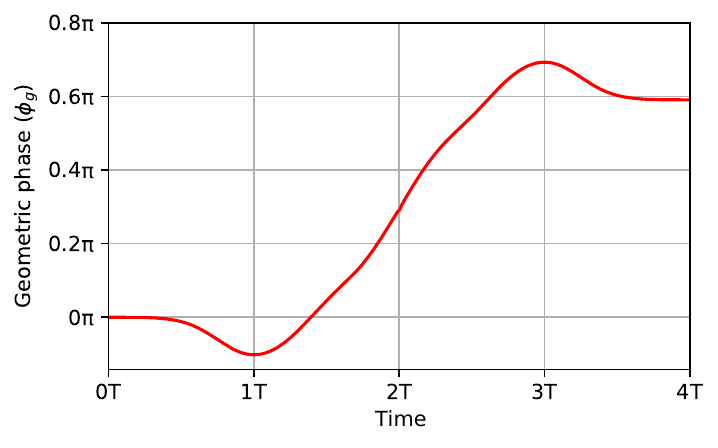}
        \caption{GP acquired by $\ket{00}$ state} during MS interaction as a function of time.
    \label{fig:ff_GP}
\end{figure}

\noindent \textbf{SF Regime}\\
The trajectory of the quantum state in the SF regime is significantly more intricate compared to the WF regime, as evident from the population dynamics of all four energy levels shown in Fig.~\ref{fig:ff}. This non-trivial trajectory naturally leads to a nonzero GP, which is plotted as a function of time in Fig.~\ref{fig:ff_GP}. 
\begin{figure}[!h]
    \centering
\includegraphics[width=0.95\linewidth]{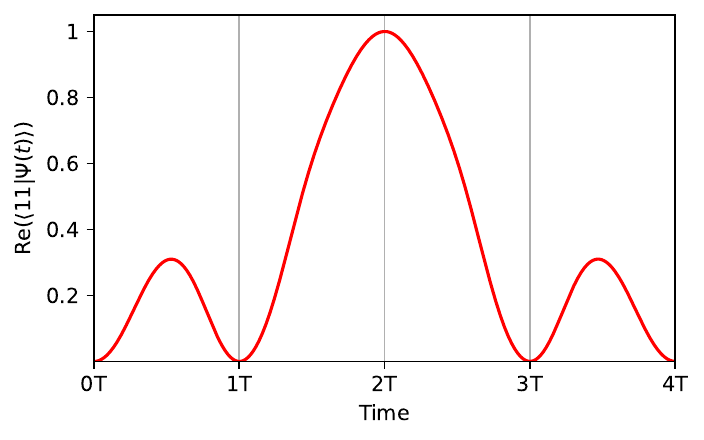}
    \caption{Real part of $\braket{11}{{\Psi(t)}}$} indicating sharp change around $2T$
    \label{fig:slope_re}
\end{figure}
\begin{figure}[!h]
    \centering
    \includegraphics[width=0.95\linewidth]{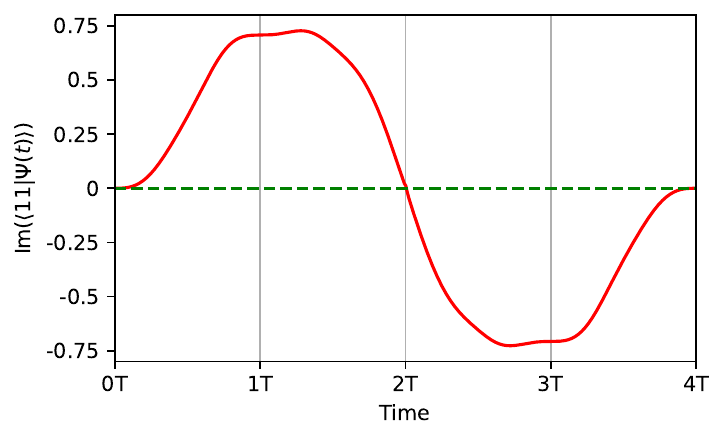}
    \caption{Imaginary part of $\braket{11}{{\Psi(t)}}$} indicating sharp change around $2T$
    \label{fig:slope_im}
\end{figure}

Using Eq.~\eqref{gp_expr}, we observe that the global phase in the SF regime remains zero at all times because the inner product \( \braket{\Psi(0)}{\Psi(t)} \) is always real. On the other hand, the dynamic phase is nonzero whose behavior can be inferred from the eigenvector $\ket{\Psi(t)}$. 

The first point to be noted is the probability amplitudes $\braket{10}{\Psi(t)} \text{ and } \braket{01}{\Psi(t)}$ are zero for all values of $t$. Further $\braket{00}{\Psi(t)}$ is always real. Thus the only probability amplitude of $\ket{\Psi(t)}$ which contributes to a dynamical phase is $\braket{11}{\Psi(t)}$. Both the real and imaginary parts of $\braket{11}{\Psi(t)}$ have been plotted in Fig: \ref{fig:slope_re} and Fig: \ref{fig:slope_im} respectively. At odd clock pulse times i.e. at $1T$ and $3T$, there is change in direction of GP as seen in Fig: \ref{fig:ff_GP}. This can be directly inferred from the change in slope in both real and imaginary parts of $\braket{11}{\Psi(t)}$. Another point to be noted is that at $2T$ there is a sharp change in $\braket{11}{\Psi(t)}$, particularly in the imaginary part. While this sharp change does not contribute to any significant deviations in GP picked up in during the unitary evolution, it does appear to be a \textit{sensitive zone}. When the quantum state undergoes non-unitary evolution this zone is where one might expect significant deviations. 

\section{Environmental interaction in SF regime}
\begin{figure*}[t]
    \centering
    \begin{tabular}{c c}
   \includegraphics[width=0.9\columnwidth]{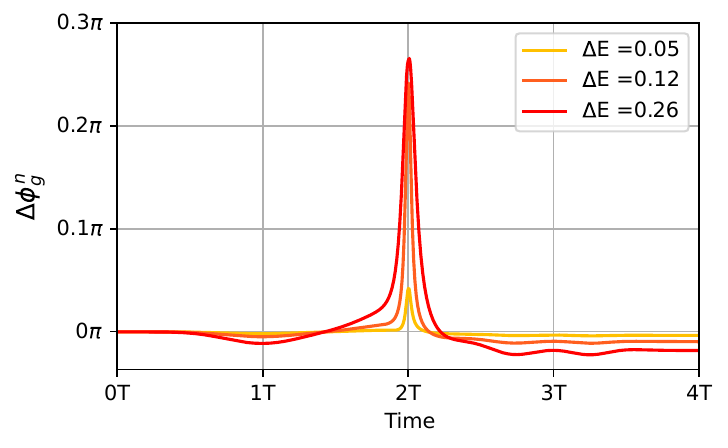} 
   
   &\includegraphics[width=0.9\columnwidth]{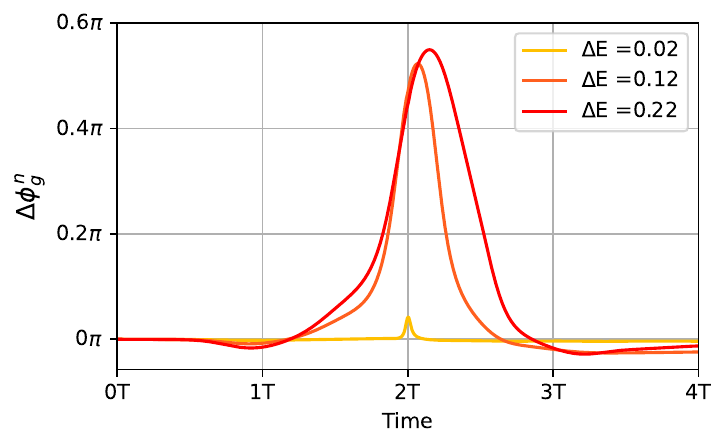}  \\
   Qubit dephasing & Qubit decay\\
     \includegraphics[width=0.9\columnwidth]{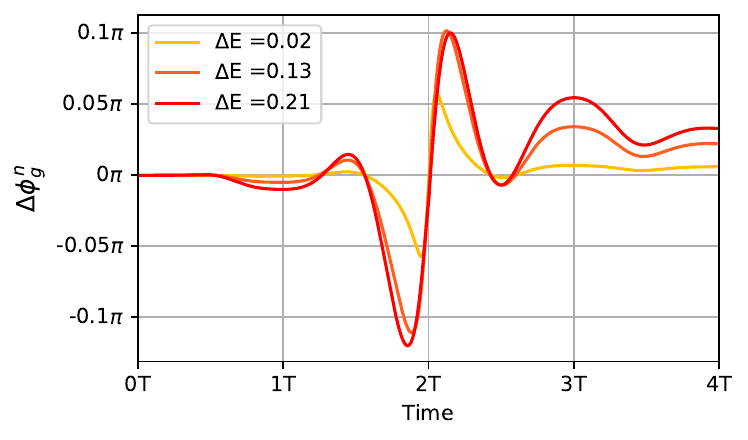}&\includegraphics[width=0.9\columnwidth]{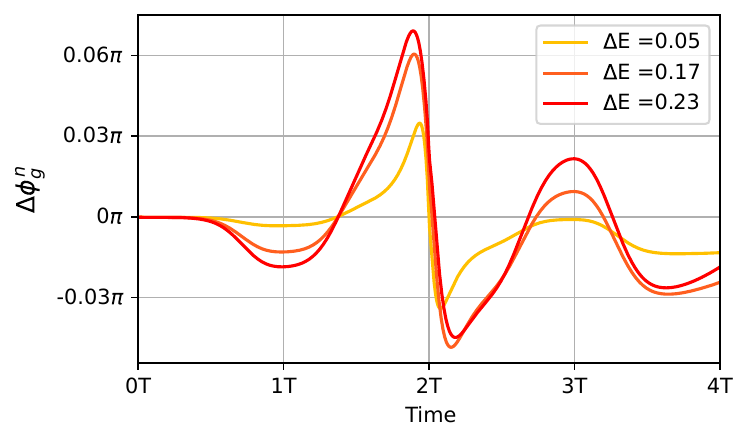} \\
     Motional heating & Motional dephasing
\end{tabular}
    \caption{GP acquired by the bipartite system when the MS interaction is plagued by common noise models mentioned in Table:~\ref{tab:my_label} causing different values of entanglement loss ($\Delta E$).}
    \label{fig:gp-deviation}
\end{figure*}
As discussed previously, this GP could be affected when the system is prone to environmental noise. 
The density matrix $\rho(t)$ of a quantum state undergoing such non-unitary dynamics is given by the Lindblad master equation\cite{Thetheoryofopenquantumsystems,ogLindblad,oglindblad2} which is expressed in the canonical form as 
\begin{equation}
i\hbar\frac{\partial\rho}{\partial t} =\left[H_I, \rho\right]+\sum_k \gamma_k\left(L_k \rho L_k^{\dagger}-\frac{1}{2} L_k^{\dagger} L_k \rho-\frac{1}{2} \rho L_k^{\dagger} L_k\right).
\end{equation}
In this work, we have considered the  four most commonly encountered environmental noises in trapped ion systems which have been tabulated in Table~\ref{tab:my_label}~\cite{thesis,noise_list}.
\begin{table}[H]
    \centering
    \begin{tabular}{|p{0.2\columnwidth}|p{0.2\columnwidth}|p{0.40\columnwidth}|}\hline 
        \centering Error type &\centering Lindbladian &Causes  \\ \hline 
        \centering Qubit Decay&\centering$\hat{\sigma}_{-}$ &-Spontaneous emission noise\newline-Magnetic field noise\\ \cline{1-2}
        \centering Qubit Dephasing&\centering$\hat{\sigma}_z$ &-Drive frequency noise \\ \hline 
        \centering Motional heating&\centering$\hat{a}$ &-Electric field noise\newline-Spontaneous emission\newline\hspace*{0.08ex} on phonon sideband\\ \cline{1-2}
        \centering Motional dephasing&\centering$\hat{a}^{\dagger}\hat{a}$ &-Motional frequency\newline\hspace*{0.08ex} fluctuations\\ \hline 
    \end{tabular}
    \caption{Common noises in Trapped-ion systems}
    \label{tab:my_label}
\end{table}
 \begin{figure}[h]
    \centering
   \subfigure{\includegraphics[width=1\columnwidth]{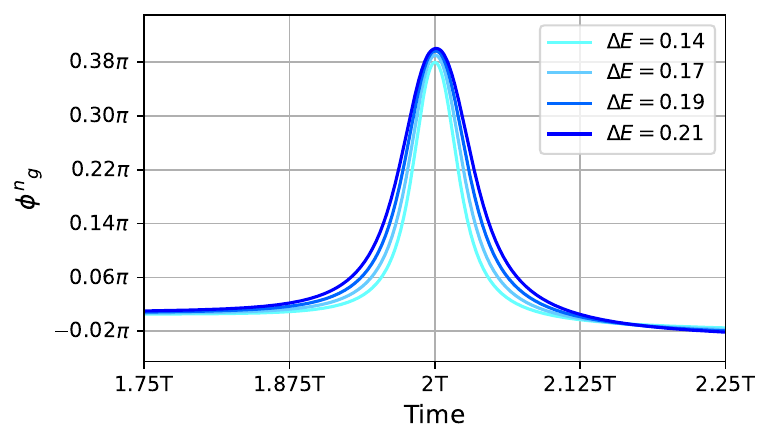}}
   \subfigure{\includegraphics[width=0.95\columnwidth]{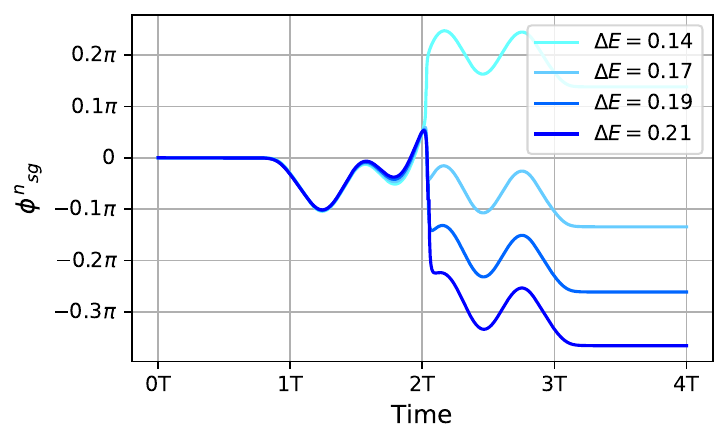}}
    \caption{GP acquired by the bipartite system and subsystem $\rho_B$ when the MS gate interaction is plagued by a non-local noise $L_{nl}$ causing different values of entanglement loss ($\Delta E$). }
    \label{fig:gp-subsys}
\end{figure}
 The deviation of the GP under the influence of these interactions has been plotted in Fig:~\ref{fig:gp-deviation}. Change in GP is represented as $\Delta\phi_g(t) = \phi_g^{u}(t) - \phi_g^{n}(t)$, where $\phi_g^{u}(t)$ and $\phi_g^{n}(t)$ represents GP acquired at time $t$ during unitary and non-unitary evolution respectively. The change in entanglement due to these noises is expressed as 
 \begin{equation}
     \Delta E = 1 - E(T)
 \end{equation}
 where $E(T)$ is the entanglement of the quantum state measured using negativity at gate pulse time $T$.\\
 \textbf{Discussion}\\
In all four plots, a sharp behavior of \( \Delta\phi_g \) is observed at \( 2T \), apart from mild deviations at other times. This can be directly inferred from the fact that the rate of change of \( \ket{\Psi(t)} \) is significantly pronounced around \( 2T \), as shown in \cref{fig:slope_re,fig:slope_im}. While the presence of noise may slightly alter \( \ket{\Psi(t)} \) and \( \ket{\dot{\Psi}(t)} \), these modifications can have a considerable impact on the inner product \( \braket{\Psi(t)}{\dot{\Psi}(t)} \). This sharp change in GP for noisy processes around \( 2T \) provides an effective means to detect both the type and strength of noise present in the system.  

\section{Subsystem dynamics}
So far we have only discussed the GP acquired by the two qubit quantum state. In this section, we discuss about the GPs picked up by the subsystem during the MS interaction. All two-qubit density operators which are of the form:
\begin{equation}
    \rho = \begin{bmatrix}
        a & 0 & 0 & w\\
        0 & b & z & 0\\
        0 & z^{*} & c & 0\\
        w^{*} & 0 & 0 & d
    \end{bmatrix}
\end{equation}
are known as x-states~\cite{x-type}. These x-states get their name from the non-zero components of the density operator and could be generalized to any finite-dimensional quantum system. 
Since the density operator under the MS interaction is always an x-state with the initial state being pure, the GP picked up by subsystems is zero at all times. The noises discussed in previous sections preserve the "\textit{x-ness}" of the density operator. But there are certain noise models that do break the x-structure of the density operator. N. Quesada \textit{et al.} have provided a detailed analysis of the conditions for a noise model to break the \textit{"x-ness"} of a density operator~\cite{x-type}. Thus any non-negligible value of GP picked up by the subsystem during the MS interaction is to be considered as the presence of such noises.  An example of such a scenario where GP  acquired by the subsystem due to a non-local noise modeled by a Lindbladian $L_{nl} = {\hat{\sigma}^{(1)}}_x +{\hat{\sigma}^{(2)}}_z $ has been plotted in Fig:~\ref{fig:gp-subsys}.

\section{Conclusion}
In this work, we have studied the GP acquired by a quantum state during a MS interaction in both the WF and SF regimes. We demonstrate that GP can serve as a reliable tool to verify the accuracy of achieving the WF regime. Furthermore, we show that by analyzing the behavior of GP around \( 2T \), one can identify both the strength and type of noise affecting the MS gate implementation in the SF regime. Additionally, we have also discussed the possibility of using the GP acquired by the subsystems as a signature of certain kinds of noises.

Since the MS interaction is a powerful mechanism for generating multipartite entanglement, this study could be extended to multipartite scenarios, where the experimental feasibility of quantum-state tomography becomes more challenging. While our analysis has focused on trapped-ion systems, similar investigations could be conducted for other qubit platforms, potentially simplifying the process of noise identification during quantum gate implementations.

\section*{Acknowledgements}
D.R would like to thank Sharad Mishra for his inputs on differential geometry. We would like to thank Prof. Joseph Samuel for his discussions. The authors are grateful to Dr. Ludmila Vioti, who accepted our invitation to present her work in our department, inspiring us to explore this topic. We also thank Dr. Jay Mangaonkar and Dr. Saurabh V. Kadam for their insights into the experimental aspects of trapped-ion systems. \\
 All calculations and simulations in this work were performed using the QuTiP library \cite{qutip1,qutip2}.
 \bibliography{references.bib}
\end{document}